# Statistical study of time intervals between murders for serial killers

Enzo Yaksic
Serial Homicide Expertise and Information Sharing Collaborative, Boston, MA, USA

M.V. Simkin and V.P. Roychowdhury
Department of Electrical and Computer Engineering, University of California, Los Angeles, CA 90095-1594

**Abstract**

We study the distribution of 2,837 inter-murder intervals (cooling off periods) for 1,012 American serial killers. The distribution is smooth, following a power law in the region of 10-10,000 days. The power law cuts off where inter-murder intervals become comparable with the length of human life. Otherwise there is no other characteristic scale in the distribution. In particular, we do not see any characteristic spree-killer interval or serial-killer interval, but only a monotonous smooth distribution lacking any features. This suggests that there is only a quantitative difference between serial killers and spree-killers, representing different samples generated by the same underlying phenomenon. The over decade long inter-murder intervals are not anomalies, but rare events described by the same power-law distribution and therefore should not necessarily be looked upon with suspicion, as has been done in a recent case involving a serial killer dubbed as the "Grim Sleeper." This large-scale study supports the conclusions of a previous study, involving three prolific serial killers, and the associated neural net model, which can explain the observed power law distribution.

## Introduction

Although there is plenty of scholarly literature on serial killers, there is not much quantitative study of time intervals between murders that are also known as cooling-off periods. Researchers mostly state that they "range from days to weeks or months" (Geberth, 2006). Or compute the mean and median of these intervals (Osborn and Salfati, 2015). For a long time, the only advanced mathematical study in this field was that of Lange (1999) who attempted a catastrophe theory approach to the prediction of patterns of inter-murder intervals. In a recent study Simkin and Roychowdhury (2014) statistically investigated inter-murder intervals for three prolific serial killers. While most of the intervals were of the order of few days some intervals were months and very few were years long. They found that these intervals follow a power-law distribution (Simkin and Roychowdhury, 2011). They also proposed a stochastic neural net model which can explain the observed power law distribution of inter-murder intervals. However, that study used modest statistical data: only 96 inter-murder intervals. Meanwhile Yaksic (2015) assembled a database of almost all American serial killers. Here we repeat the previous statistical analysis using a much larger dataset of 2,837 inter-murder intervals for 1,012 serial killers. This large-scale data supports the conclusions and modeling of Simkin and Roychowdhury (2014).

## Methods

According to current FBI definition serial murder is "unlawful killing of two or more victims by the same offender(s), in separate events" (Morton & Hilts, 2008). 'Serial Homicide Expertise and Information Sharing Collaborative' and 'Radford/FGCU Serial Killer Database Project' (Yaksic, 2015) contains data for most American serial killers.

From the database, we selected lone killers, that is those who never committed a murder with an accomplice. Even if a killer committed a single murder with an accomplice and many murders on his own he was excluded at this stage.

Not for every murder the database has the exact date. Often only the month or even only the year. We selected those lone killers for which we have the exact dates for each of their murders. Even if a killer committed a single murder on an uncertain date and many murders on certain dates he was excluded at this stage.

The database does not contain exact times of murders, only the date. So, we could not study the inter-murder intervals of less than a day. Some of the killers committed all their murders on the same date (though in separate events). Such killers were excluded at this stage for they do not give us known inter-murder intervals. However, if the killer committed only some of the murders on a single date(s) he was included in the analysis. We just merged all murders committed on a single date(s) into single event(s) labeled by the date(s).

At the end, we were left with 1,012 killers who committed murders on at least two different dates. There were a total of 2,837 inter-murder intervals for those killers.

Although FBI had changed the minimum number of killing events for a serial killer from three to two (Morton & Hilts, 2008) criminologist objected to this change insisting that three killing events are necessary to call someone a serial killer (Fox and Levin, 2014). To account for this point of view we purged the killers with only two killing dates from our sample and studied 587 killers with at least three killing dates and their 2,412 inter-murder intervals.

Finally, we selected 34 real serial killers, with at least 10 killing dates, and analyzed their 607 inter-murder intervals.

We did all the operations using Microsoft Excel and Access.

## Results

We will start with the distribution of serial killers regarding the number of killing dates. To represent the data, we will use the so-called logarithmic binning which is customary in studying data that follow a power-law distribution (Simkin and Roychowdhury, 2011). To the first bin (see Table 1) go the killers with 2 kill dates. To the second - those with more than 2 but less than or equal to 4. To the third - those with more than 4 but less than or equal to 8. And so on. The upper boundary of each subsequent bin increases twice. The size of each subsequent bin also increases twice. However, on the logarithmic scale (see Figure 1) the bin boundaries are equally spaced. So comes the name. Such binning is necessary because if we use conventional binning the vast majority of bins on the upper end of the distribution will be empty. We compute the frequency distribution by dividing the number of killers in the bin by the size of the bin and dividing the result by the total number of killers. We use the observed frequency distribution as an estimate of the probability distribution (Moore and McCabe, 1993).

As one can see from Figure 1 we can well approximate the probability distribution by a power law (here $n$ is the number of killing dates)

$$p(n) = C \times n^{-\gamma} \qquad (1)$$

with $C \approx 4$ and $\gamma \approx 2.5$.

The number of killing dates coincides with the total number of killings in the case when the killer always killed only one person on a single day. Since it is almost always the case the distribution of the killers by the victim count will be almost identical to the above.

For these 1,012 killers, there were 2,837 inter-murder intervals. The longest interval is 16, 963 days which is over 46 years. Table 2 shows the distribution of their lengths. It uses the same logarithmic binning as Table 1. We do not show the lower boundary of the bins this time, only the upper bound.

We plotted the data of Table 2 in Figure 2. There is a good power-law fit for the range of inter-murder intervals between 10 and 10,000 days, given by Eq.(1) (this time $n$ is the length of intervals in days) with $C \approx 0.4$ and $\gamma \approx 1.16$. There is a drop off at high intervals, where they become comparable with the length of human life (10,000 days is 27 years). This is apparently the only characteristic scale in the problem.

There is a controversy of what constitutes a serial killer with regard to the minimum killing events count with many researchers demanding at least three events (Fox and Levin, 2014). So we selected the killers with at least 3 killing dates. There were 587 of those with the total of 2,412 inter-murder intervals. The longest interval is 11,804 days or over 32 years. Table 3 shows their distribution. We plotted the data of Table 3 in Figure 3. Again there is a good power-law fit for the range between 10 and 10,000 days. This time with slightly different parameters $C \approx 0.6$ and $\gamma \approx 1.23$.

We also selected the real serial killers, those with at least 10 kill dates. There were 34 of those and 607 inter-murder intervals. The longest is 5,673 days or over 15.5 years. The distribution is in Table 4. We plotted the data of Table 4 in Figure 4. Once more there is a good power-law fit for the range between 10 and 10,000 days. Again with different parameters $C \approx 1.7$ and $\gamma \approx 1.46$
.

**Table 1**. Distribution of 1,012 killers by the number of kill dates.

| Number of kill dates | Number of killers | Probability |
|---|---|---|
| 2 | 425 | 0.41996 |
| 3-4 | 371 | 0.1833 |
| 5-8 | 171 | 0.042243 |
| 9-16 | 31 | 0.003829 |
| 17-32 | 10 | 0.000618 |
| 33-64 | 4 | 0.000124 |

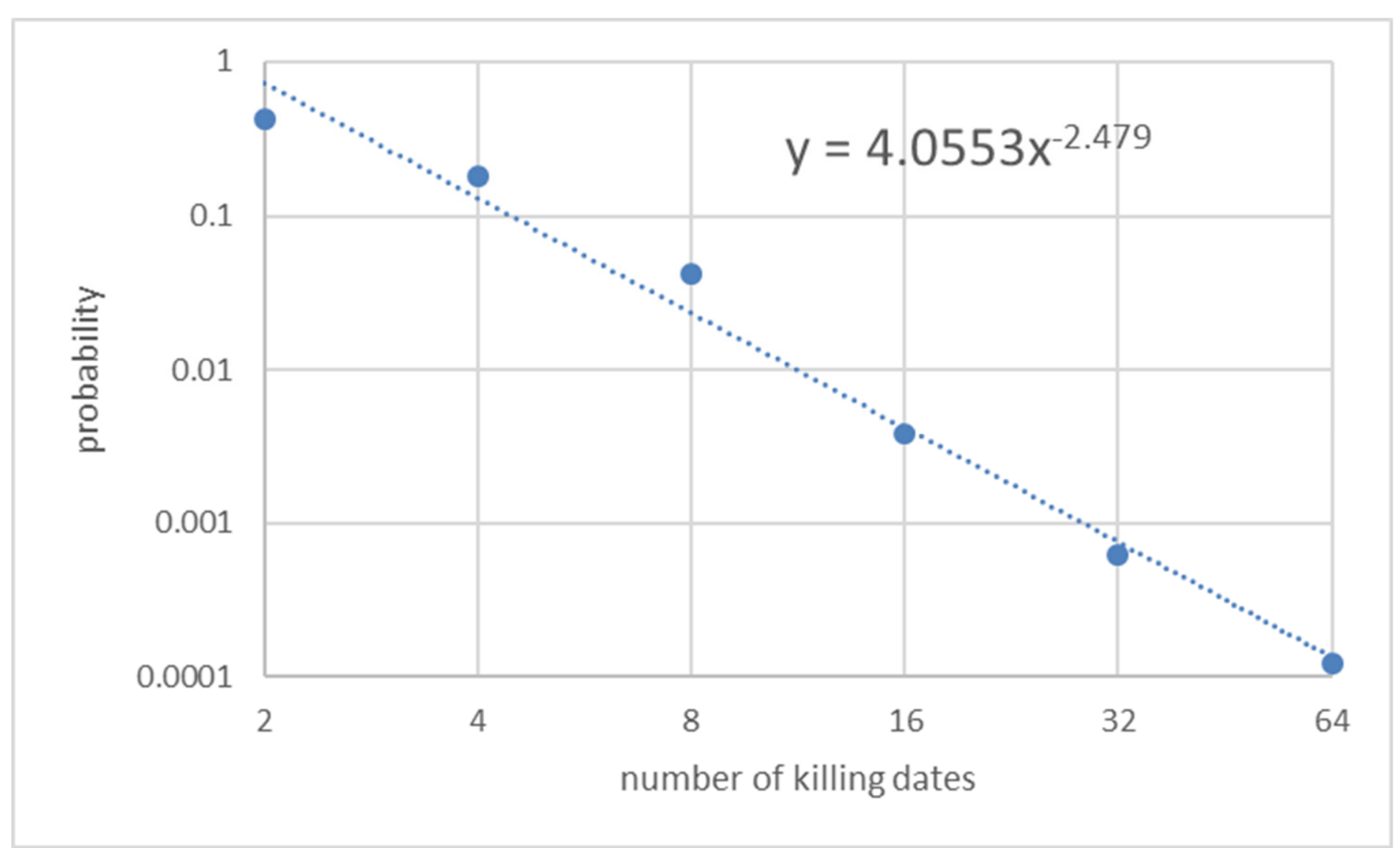

**Figure 1.** The distribution of 1,012 killers by the number of killing dates.

**Table 2.** Distribution of length of 2,837 inter-murder intervals for 1,012 serial killers.

| Upper Boundary of Intervals (in days) | Number of intervals | Probability |
|---|---|---|
| 1 | 111 | 0.039126 |
| 2 | 67 | 0.023616 |
| 4 | 127 | 0.022383 |
| 8 | 170 | 0.014981 |
| 16 | 275 | 0.012117 |
| 32 | 321 | 0.007072 |
| 64 | 290 | 0.003194 |
| 128 | 284 | 0.001564 |
| 256 | 286 | 0.000788 |
| 512 | 264 | 0.000364 |
| 1024 | 196 | 0.000135 |
| 2048 | 192 | 6.61E-05 |
| 4096 | 140 | 2.41E-05 |
| 8192 | 94 | 8.09E-06 |
| 16384 | 19 | 8.18E-07 |
| 32768 | 1 | 2.15E-08 |

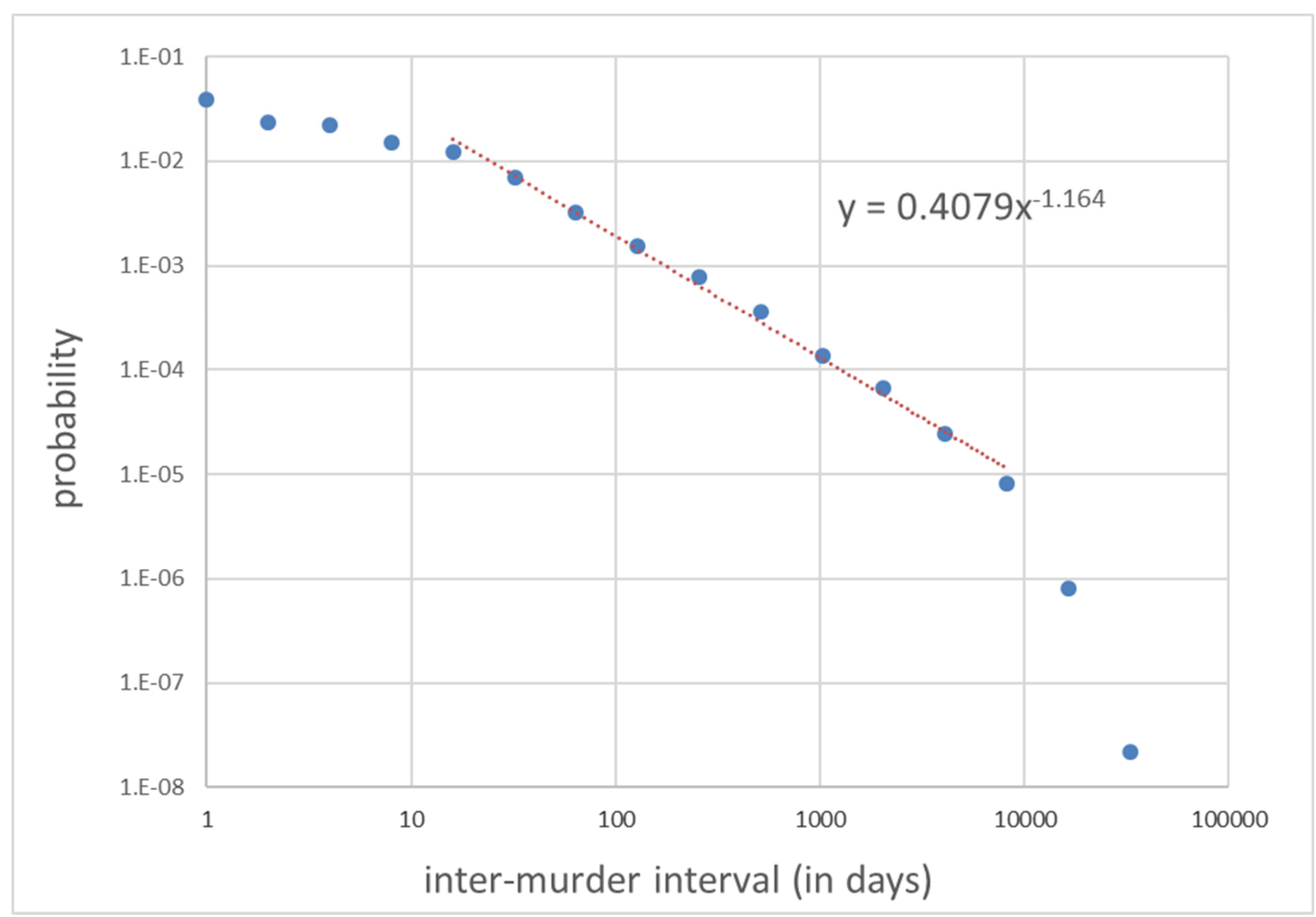

**Figure 2.** Distribution of length of 2,837 inter-murder intervals for 1,012 serial killers (circles). The line is a power-law fit.

**Table 3.** Distribution of length of 2,412 inter-murder intervals for 587 serial killers who had at least 3 killing dates.

| Upper Boundary of Intervals (in days) | Number of intervals | Probability |
|---:|---:|---:|
| 1 | 92 | 0.038143 |
| 2 | 59 | 0.024461 |
| 4 | 114 | 0.023632 |
| 8 | 148 | 0.01534 |
| 16 | 245 | 0.012697 |
| 32 | 292 | 0.007566 |
| 64 | 260 | 0.003369 |
| 128 | 256 | 0.001658 |
| 256 | 252 | 0.000816 |
| 512 | 233 | 0.000377 |
| 1024 | 172 | 0.000139 |
| 2048 | 143 | 5.79E-05 |
| 4096 | 88 | 1.78E-05 |
| 8192 | 51 | 5.16E-06 |
| 16384 | 7 | 3.54E-07 |

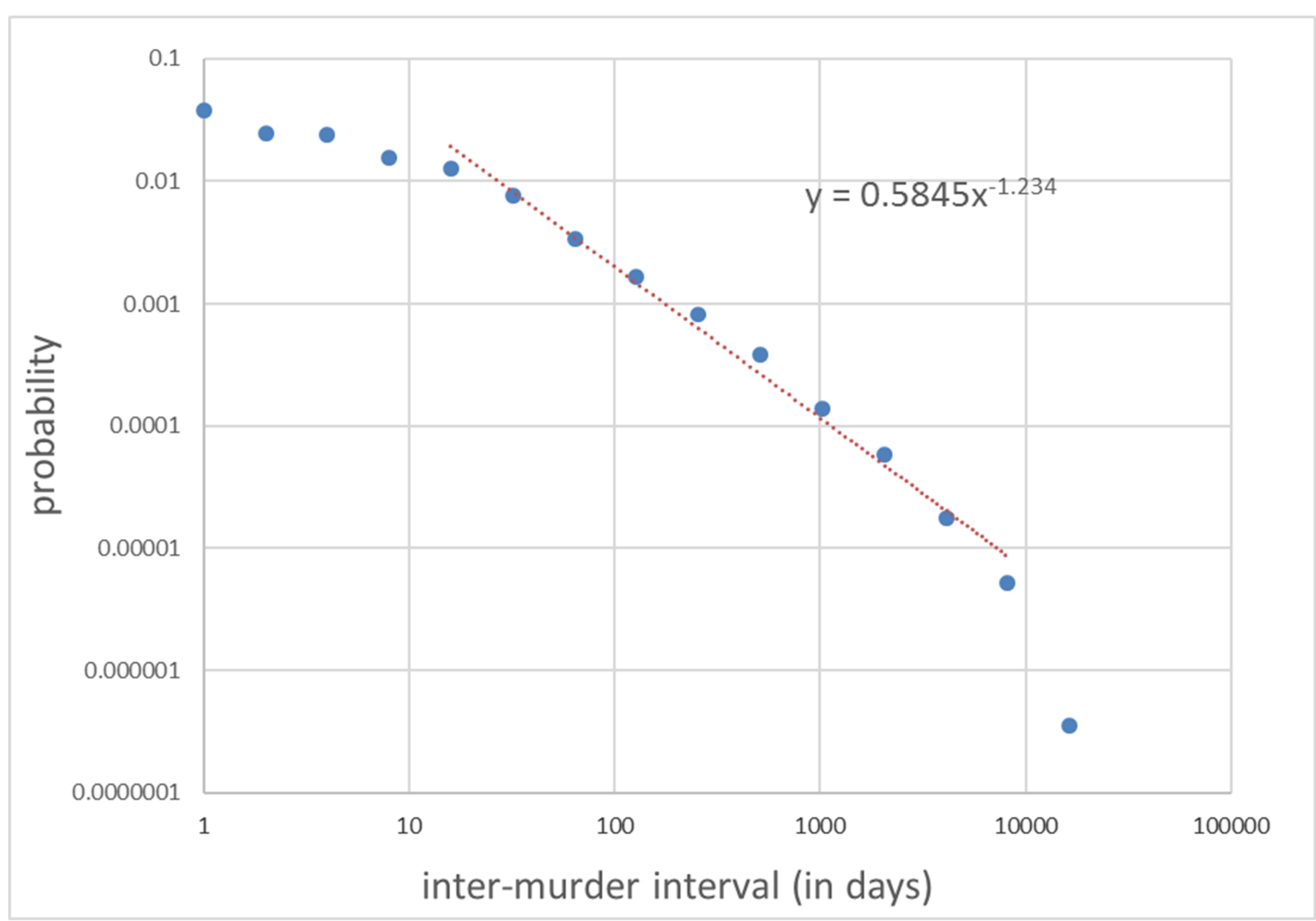

**Figure 3.** Distribution of length of 2,412 inter-murder intervals for 587 serial killers who had at least 3 killing dates (circles). The line is a power-law fit.

**Table 4.** Distribution of length of 607 inter-murder intervals for 34 serial killers who had at least 10 killing dates.

| Upper Boundary of Intervals (in days) | Number of intervals | Probability |
|---:|---:|---:|
| 1 | 21 | 0.034596 |
| 2 | 14 | 0.023064 |
| 4 | 30 | 0.024712 |
| 8 | 53 | 0.021829 |
| 16 | 74 | 0.015239 |
| 32 | 97 | 0.009988 |
| 64 | 93 | 0.004788 |
| 128 | 72 | 0.001853 |
| 256 | 57 | 0.000734 |
| 512 | 38 | 0.000245 |
| 1024 | 29 | 9.33E-05 |
| 2048 | 14 | 2.25E-05 |
| 4096 | 10 | 8.04E-06 |
| 8192 | 5 | 2.01E-06 |

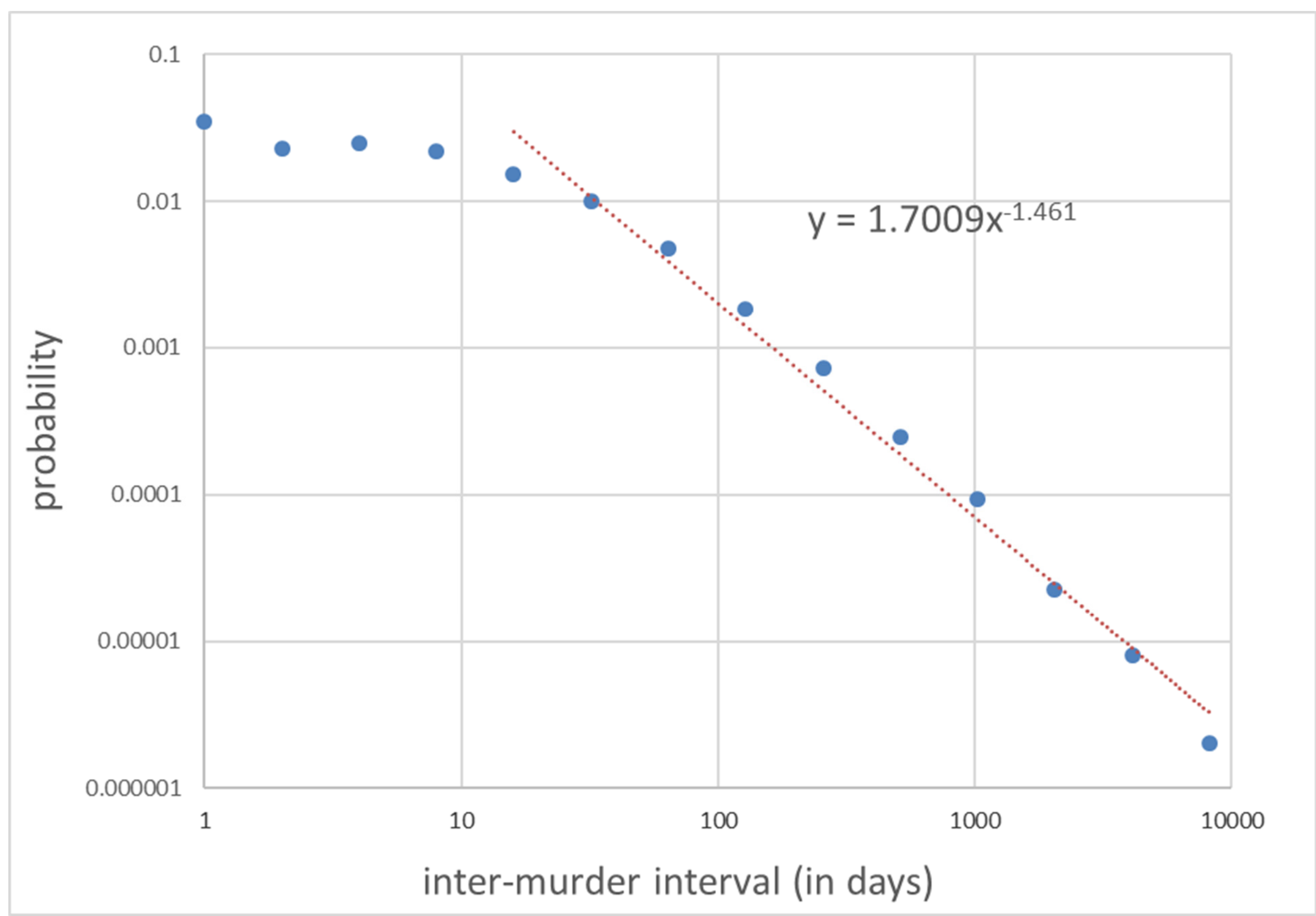

**Figure 4.** Distribution of length of 607 inter-murder intervals for 34 serial killers who had at least 10 killing dates (circles). The line is a least-square power-law fit. As described in the Discussion section and in the Appendix, the maximum likelihood estimates of the power-law exponents match the corresponding values obtained from least-square fits. There is no statistically significant discrepancy in the estimates obtained by the two methods.

## **Discussion**

As one can see from the Figures 2-4 the probability distribution of time intervals between murders is a smooth monotonously decreasing function of interval length. Note that the extremely large intervals are not anomalies but rare events governed by the same probability distribution which also describes shorter inter-murder intervals. So one should not look upon the long inter-murder intervals with suspicion. *This cautionary guidance is particularly pertinent to the case of Lonnie Franklin Jr., a serial killer nick-named as the "Grim Sleeper,"* where a gap of 13 years (1989-2002) over which no recorded murder could be attributed to him was viewed as more of a forensic failure rather than a natural outcome. The popular sentiment being that he must have murdered several victims during this so-called dormant period (Zupello (2016)). *Our study shows that such a long gap is statistically consistent and barring further evidence is not anomalous.*

The 1.46 power law exponent obtained for the killers with at least 10 killing dates is only slightly below the theoretical value of 1.5 produced by the stochastic neural net model of a serial killer (Simkin and Roychowdhury, 2014). We obtained the 1.46 exponent by least-square fitting the binned data starting with the 16-day bin. If instead we fit the data starting with the 32-day bin we get $\gamma \approx 1.54$ which is slightly above the theoretical value. A maximum likelihood estimate for the intervals of 9 or more days (since the 16-day bin contains all intervals between 9 and 16 this is a match to the least-square fit starting with the 16-day bin) gives $\gamma \approx 1.48$. A maximum likelihood estimate for the intervals of 17 or more days (this is a match to the least-square fit starting with the 32-day bin) gives $\gamma \approx 1.56$. *Both maximum likelihood estimates are very close to the corresponding least-square estimates and to the theoretical value.*

Shalizi (2012) criticized our original paper claiming that using a complementary cumulative distribution function one can show that the distribution of inter-murder intervals is not a power law but lognormal. These criticisms got widely spread over the internet and got a hold of scientific community. We therefore must refute these misunderstandings. Since this discussion will not be interesting to every reader we will do this in the Appendix.

The power law exponent does match the theoretical one for the killers with at least 10 killing dates. However, when we decrease the threshold to at least 2 killing dates the power law exponent, $\gamma$, drops to 1.16. To understand what is going on let us look at the shown in Figure 5 distribution of inter-murder intervals for 425 killers with exactly 2 killing dates.

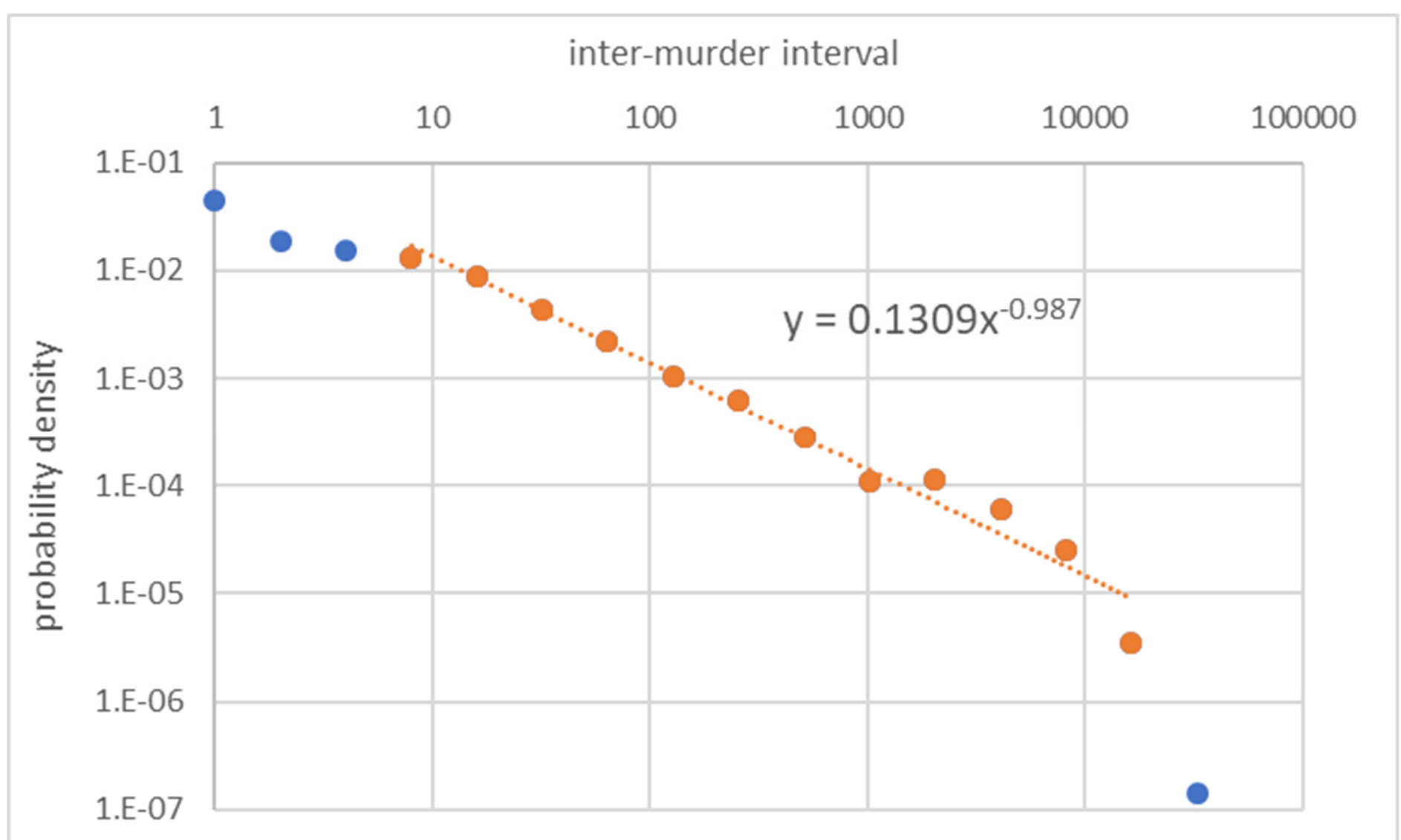

**Figure 5.** Distribution of length of 425 inter-murder intervals for 425 serial killers who had exactly 2 killing dates (circles). The line is a least-square power-law fit.

We see a power-law fit in the region 10-10,000 days with an even smaller exponent $\sim 1$. A possible explanation of this discrepancy is that a large fraction of the killers with small number of murders are not like the majority of the serial killers who plan a murder when the urge to kill crosses a threshold, as posited in (Simkin and Roychowdhury, 2014). For such serial killers the theory predicted that the urge to kill becomes irresistible with intervals that follow a power law distribution with exponent of 1.5. The other killers may be driven by other reasons. As a result, the inter-murder interval distribution may also be different. The simplest model is the Poisson process where every day there is a fixed small probability to commit a murder. This leads to an exponential distribution of inter-murder intervals. We assume that the killers with exactly two kill dates are drawn from a mixture of serial killers with a power-law distribution of inter-murder intervals, and those with exponential inter-murder distributions. In fact, using manual inspection and being guided by intuition, we were able to decompose the killers into two groups (see Table 5). Distribution A which includes 239 killers is approximately a power law with exponent 1.5.

Distribution B which includes 186 killers is approximately exponential, with $\frac{1}{5000}$ murder probability on any given day. Distributions A and B are plotted in Figure 6. This partitioning can potentially be done in an automated manner using a mixture model and Maximum Likelihood estimation techniques: Each serial killer's intervals are drawn either from a power law distribution with an unknown exponent or an exponential distribution with an unknown mean. Both the exponent and the mean can be estimated, as well as the most likely assignment of each killer's intervals to one of the distributions, by maximizing the likelihood of the data.

**Table 5**. Distribution of length of 425 inter-murder intervals for 425 serial killers who had exactly 2 killing dates. We decomposed the distribution in two parts: Distribution A, which is approximately a power law with exponent 1.5, and Distribution B which is approximately exponential.

| Upper boundary of intervals (in days) | Number of intervals | Probability density | Distribution A | | Distribution B | |
|---|---|---|---|---|---|---|
| | | | Number of intervals | Probability density | Number of intervals | Probability density |
| 1 | 19 | 4.47E-02 | 19 | 7.95E-02 | | |
| 2 | 8 | 1.88E-02 | 8 | 3.35E-02 | | |
| 4 | 13 | 1.53E-02 | 13 | 2.72E-02 | | |
| 8 | 22 | 1.29E-02 | 22 | 2.30E-02 | | |
| 16 | 30 | 8.82E-03 | 30 | 1.57E-02 | | |
| 32 | 29 | 4.26E-03 | 28 | 7.32E-03 | 1 | 3.36E-04 |
| 64 | 30 | 2.21E-03 | 28 | 3.66E-03 | 2 | 3.36E-04 |
| 128 | 28 | 1.03E-03 | 24 | 1.57E-03 | 4 | 3.36E-04 |
| 256 | 34 | 6.25E-04 | 26 | 8.50E-04 | 8 | 3.36E-04 |
| 512 | 31 | 2.85E-04 | 19 | 3.11E-04 | 12 | 2.52E-04 |
| 1024 | 24 | 1.10E-04 | 6 | 4.90E-05 | 18 | 1.89E-04 |
| 2048 | 49 | 1.13E-04 | 11 | 4.49E-05 | 38 | 2.00E-04 |
| 4096 | 52 | 5.97E-05 | 4 | 8.17E-06 | 48 | 1.26E-04 |
| 8192 | 43 | 2.47E-05 | 1 | 1.02E-06 | 42 | 5.51E-05 |
| 16384 | 12 | 3.45E-06 | | | 12 | 7.88E-06 |
| 32768 | 1 | 1.44E-07 | | | 1 | 3.28E-07 |

The major deviation of the data from the theoretical model is the flattening of the distribution at small inter-murder intervals. In our original paper (Simkin and Roychowdhury, 2014) we argued that the model predicts how often the killer will have an urge to kill. However, the killer may not have an opportunity to do this. Especially an opportunity to do this without taking a high risk of being caught. The most accomplished serial killers are very cautious. Krivich & Ol'gin (1993) describe how the serial killer used in our original study would often go for a hunt and return unsuccessful. This makes short inter-murder intervals less frequent than what the theory would predict.

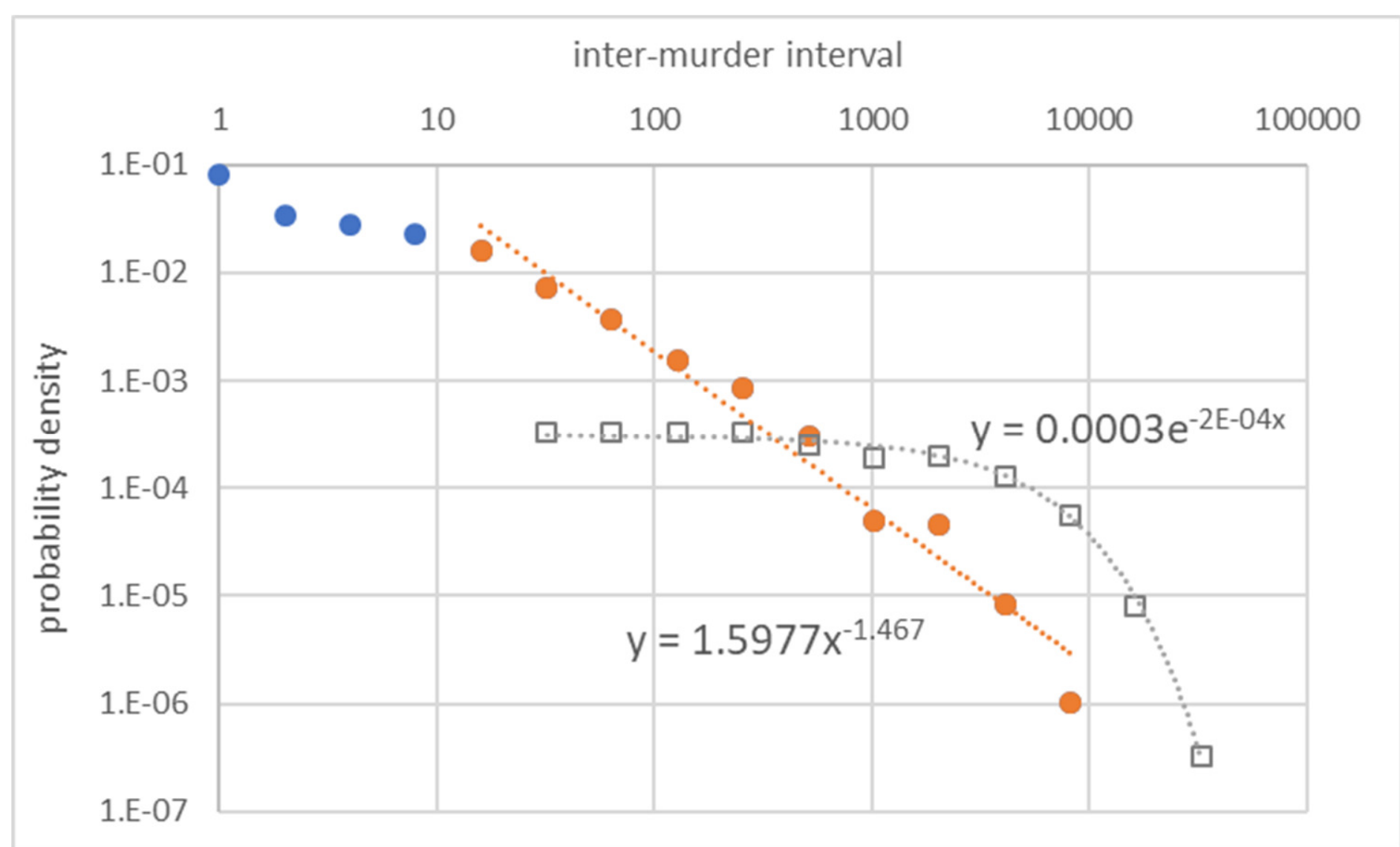

**Figure 6.** Decomposition of the distribution shown in Figure 5. Distribution A, which includes 239 killers, is shown by solid circles. Distribution B, which includes 186 killers, is shown by empty squares. Lines are least square fits.

Edelstein (2020) criticized the preprint of the present article stating :"The researchers also found that the time intervals between murders were smooth with no profound peaks of shorter or longer intervals. This contradicted an earlier study that claimed that as killers escalate their lethal behavior, so the interval between the murders gets shorter" This criticism is based on a misunderstanding. We studied the probability density function of all intervals for all killers lumped together. This function is indeed smooth with no peaks as one can see from e.g. Fig.3. We did not study the differences between the inter-murder intervals for a given killer based on their place in the sequence thus far.

Now we decided to look into this issue. We compared the first and the last inter-murder intervals for 587 killers who had at least two intervals, that is committed three or more murders. In 204 cases the last interval was longer than the first. In 14 cases they were equal. In 369 cases the last interval was shorter. So, in 63% of cases the last interval was longer. Since the standard error is only 2% the escalation effect obviously exists. Albeit we observe that the mean length of the first interval (840 days) is only 1.8 times longer than that of the last interval (469 days). Compare to the range of the intervals, which is from 1 to over 10,000 days the acceleration effect is only a small perturbation. One may argue that the sample is dominated by the consecutive intervals for the serial killers with just three murders. So, we repeated this analysis for 34 serial killers with at least 10 killing dates. For 13 of those the last interval was longer than the first one. For one killer they were equal, and for 20 the last interval was longer than the first. So, the last interval is bigger than the first in 59% of the cases. The standard error is 8% so the result is consistent with the one for the previous sample.

Some criminologists distinguish a spree killer as a separate category from a serial killer. For example, Holmes and Holmes (2010) define spree murder as "the killing of three or more people within a 30-day period" and serial murder as "the killing of three or more people over a period of more than 30 days, with a significant cooling-off period between the killings." A problem with such definition becomes evident when we look at the murder pattern of any accomplished serial killer. Figure 7 shows cumulative number of murders as a function of time for Charles Cullen. One can clearly see spree like periods when the cumulative number grows steeply and periods with large intervals between murders. For example, in 1996 Cullen killed on 5/31, 6/9, and 6/24 what makes him a spree killer according to Holmes and Holmes (2010) definition. Since 7/10/1996 until 6/22/2001 he murdered five people with the minimum inter-murder interval of over 200 days. This makes him a serial killer according to Holmes and Holmes (2010) definition.

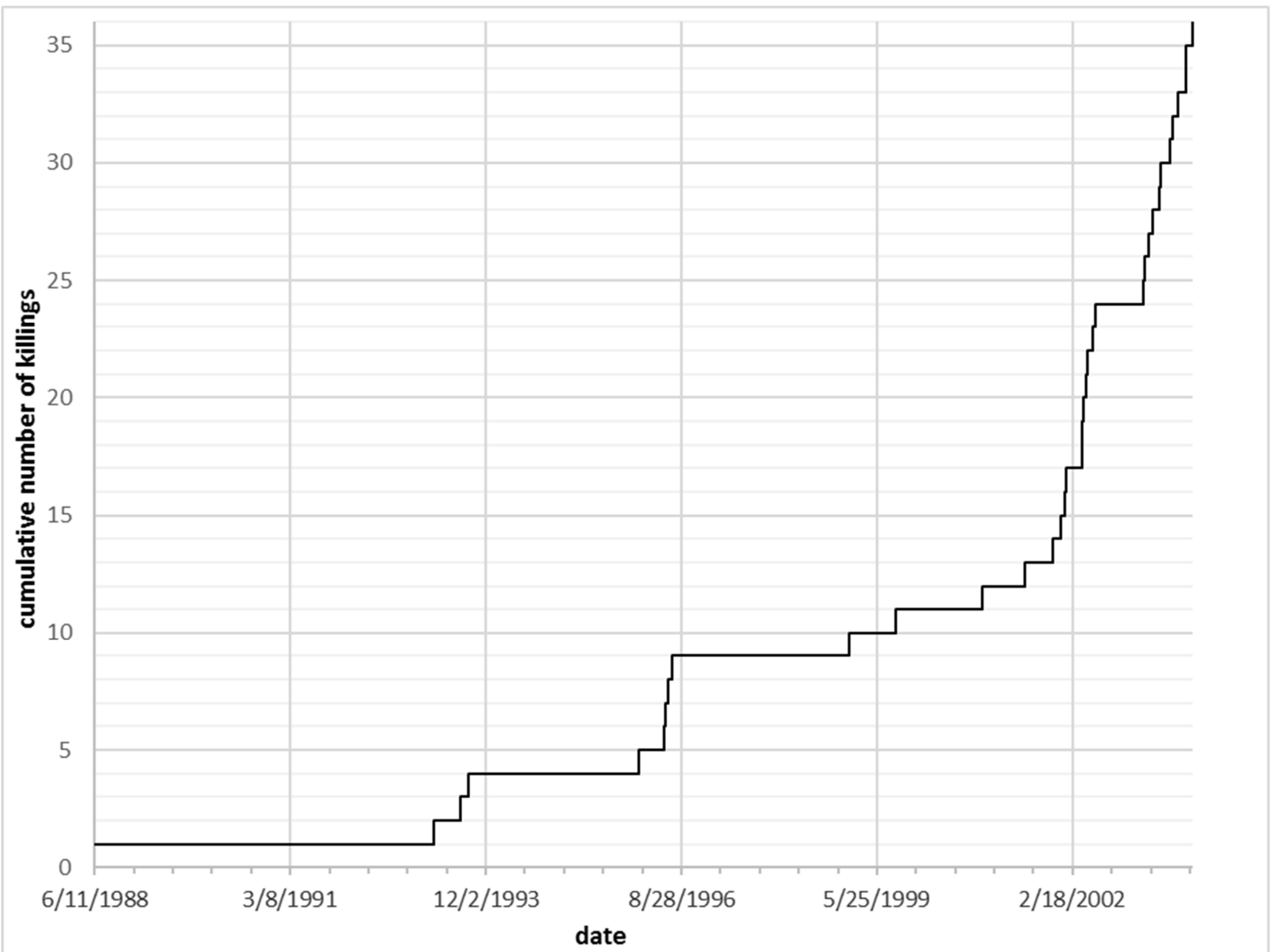

**Figure 7.** Cumulative number of killings committed by Charles Cullen. The major marks on the horizontal axis are separated by 1,000 days and the minor ones by 200 days.

FBI abandoned spree killer as a separate classification (Morton & Hilts, 2008). However, Schlesinger, Ramirez, Tusa, Jarvis, and Erdberg (2017) re-introduced this notion under the name "rapid-sequence offenders." They investigated inter-murder intervals for 44 serial killers. They found that 25 of them (56.8%) had all inter-murder intervals longer than 14 days. Six offenders (13.6%) had all intervals under 14 days. Thirteen offenders (29.5%) had some intervals shorter and some longer than 14 days. They classed the killers with inter-murder intervals under 14 days as "rapid-sequence offenders."

To understand the meaning of their results let us note that out of 2,837 inter-murder intervals in our database 76.6% are of 14 days or more. This means that if we select *m* intervals at random the probability that all of them will be 14 days or more is $0.766^m$ and the probability that all of them will be less than 14 days $0.234^m$. Now using the distribution of the 1,012 killers in our database by the numbers of murders we can compute the expected percentage of killers with all intervals of 14 days or more: $x = \frac{1}{1012}\sum N(n)0.766^{n-1}$. Here $N(n)$ is the number of killers with *n* murders. The computation gives $x = 56.9\%$ which is remarkably close to the 56.8% figure of Schlesinger et al (2017). For the expected percentage of killers with all intervals under 14 days a similar calculation gives 15.7%. This is a bit bigger than the 13.6% figure of Schlesinger et al (2017) , but they got only six criminals like that. So, the difference is well within the sampling error. So, the results of Schlesinger et al (2017) can be explained by the ordinary law of chances rather than by intrinsic differences between different categories of killers. To illustrate this point let us consider a study where instead of 44 different killers we will take 44 different coins and toss each of them a few times. Some of the coins will produce all heads, some – all tails, and some – a mixture of heads and tails. One may be tempted to classify some of the coins as head-coins and tail-coins. This, however, would be wrong since one can obtain the same result with a single coin tossed 44 times.

If we look at Figures 2-4, we do not see any characteristic spree-killer interval or serial-killer interval but a monotonous smooth distribution lacking any features. This suggests that there is only a quantitative difference between serial killers and spree-killers which represent merely different aspects of the same phenomenon.

## Appendix: how cumulative distribution functions can mislead the unwise

Shalizi (2012) criticized our first paper (Simkin and Roychowdhury, 2014) claiming that using a complementary cumulative distribution function one can show that the probability density functions of inter-murder intervals are not power law but lognormal. We repeat their analysis with our new data. See Figure A1. It indeed seems that a lognormal distribution fits the data much better.

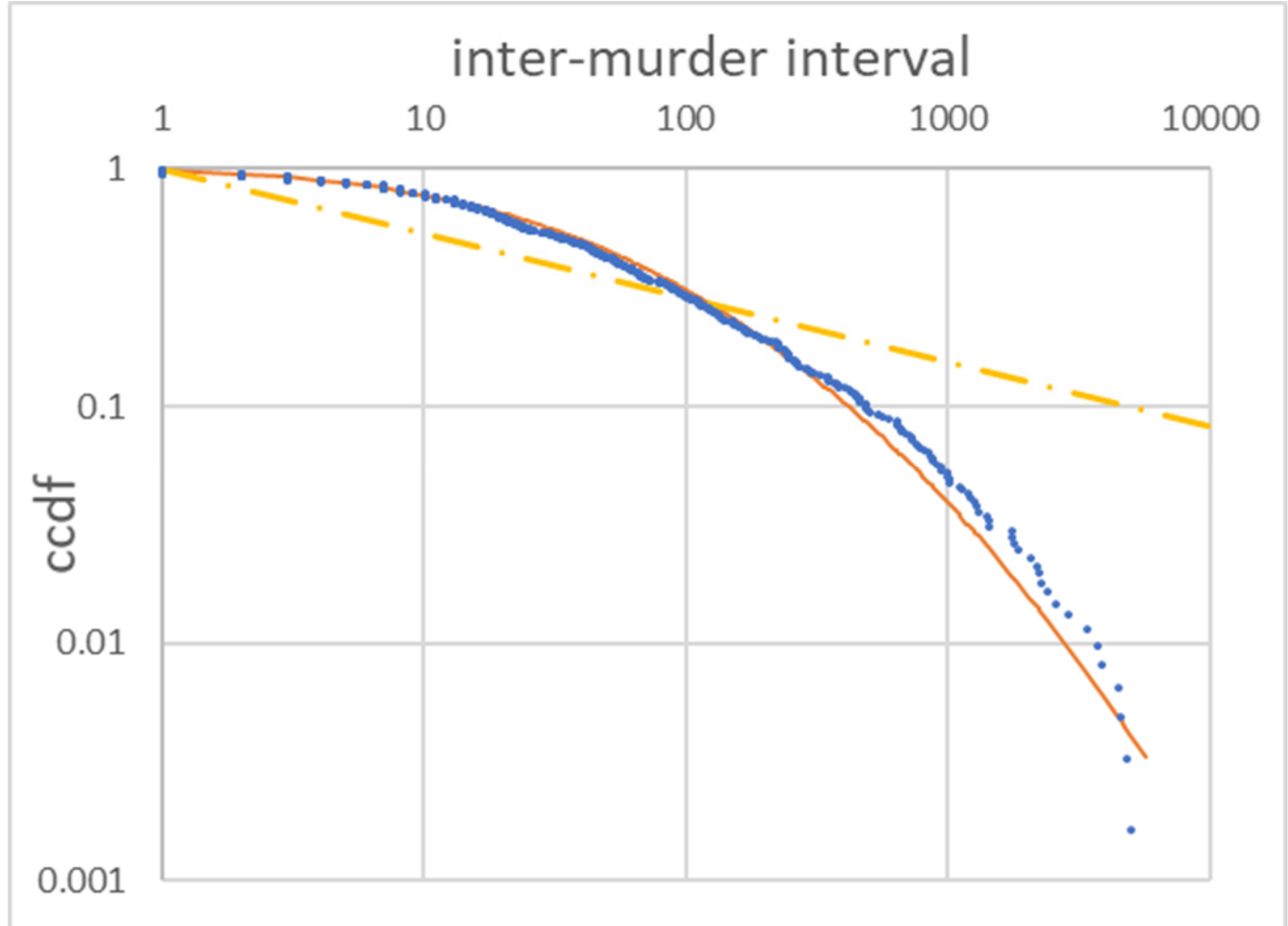

*Figure A1. Complementary cumulative distribution function of 607 inter-murder intervals for 34 serial killers who had at least 10 killing dates.* ***Dots*** *– actual data.* ***Solid line*** *– maximum likelihood lognormal fit.* ***Dot-dashed line*** *– maximum likelihood power law fit.*

On the other side if we look at the probability density function plot (Figure A2) *we see that the least square power law fit (dashed line) to the tail of the distribution is closer to the data than the lognormal fit*.

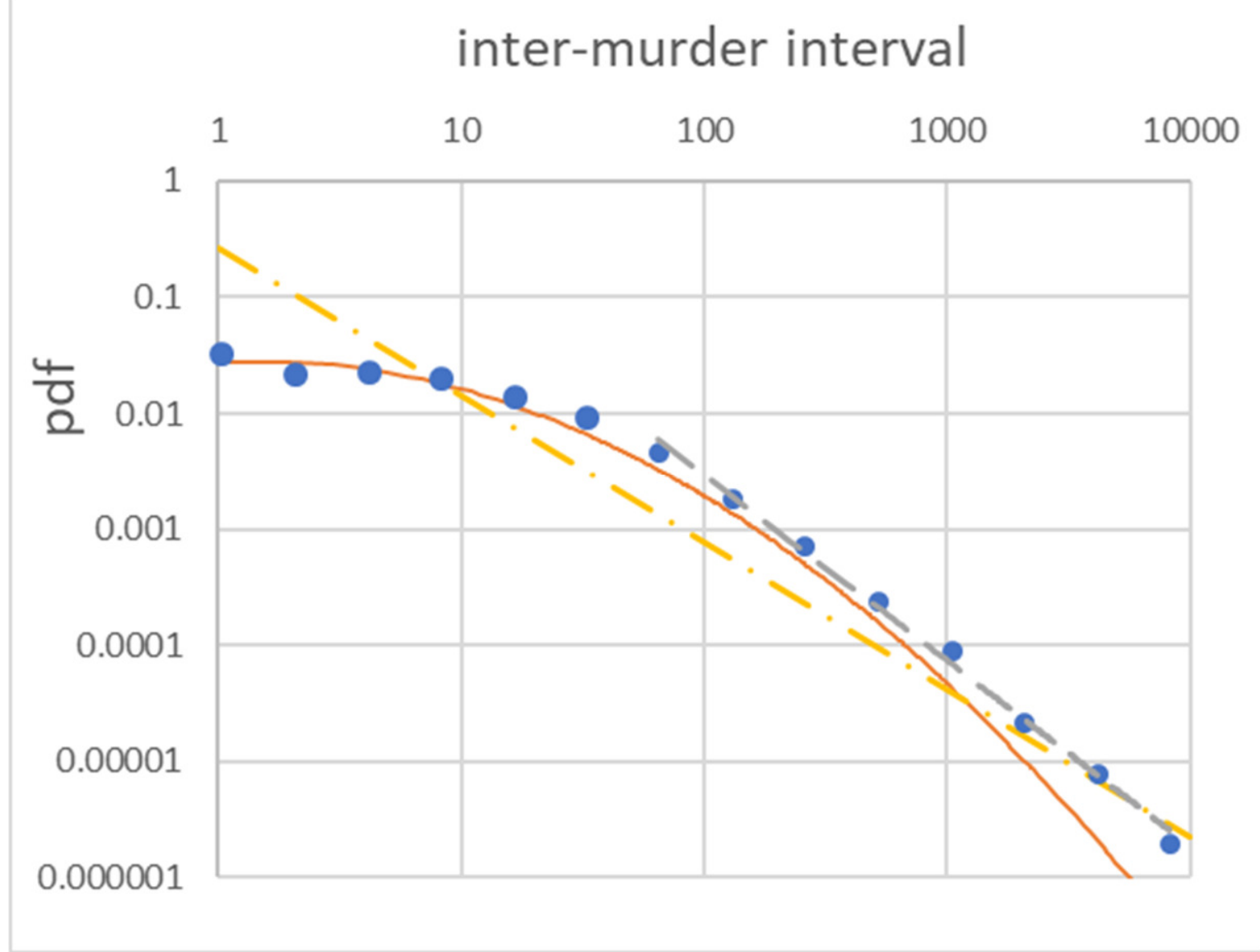

*Figure A2. Probability density function of 607 inter-murder intervals for 34 serial killers who had at least 10 killing dates.* ***Circles*** *– actual binned data.* ***Solid line*** *– maximum likelihood lognormal fit.* ***Dot-dashed line*** *– maximum likelihood power law fit.* ***Dashed line*** *– least square fit to the tail of the distribution.*

Note that the max likelihood power law fit is way off not because max likelihood produces very different result from the least square fit to the binned data, but because following our critic we did max likelihood fit to the whole data range. If we do a max likelihood fit to the tail of the distribution we get a very similar result to the least square fit. Analogously if we do a least square fit to the whole data range we get a line which is just as way off as what the max likelihood fit produced.

So, what has happened? Does CCDF indeed offer an insight into the nature of things which a PDF can't deliver? We decided to check. Two thousand data points were produced using a random number generator to follow a power law distribution $p(x) = x^{-2}$ for x > 1. In Figure A3 you see the binned distribution of these random numbers (a) and a CCDF (b).

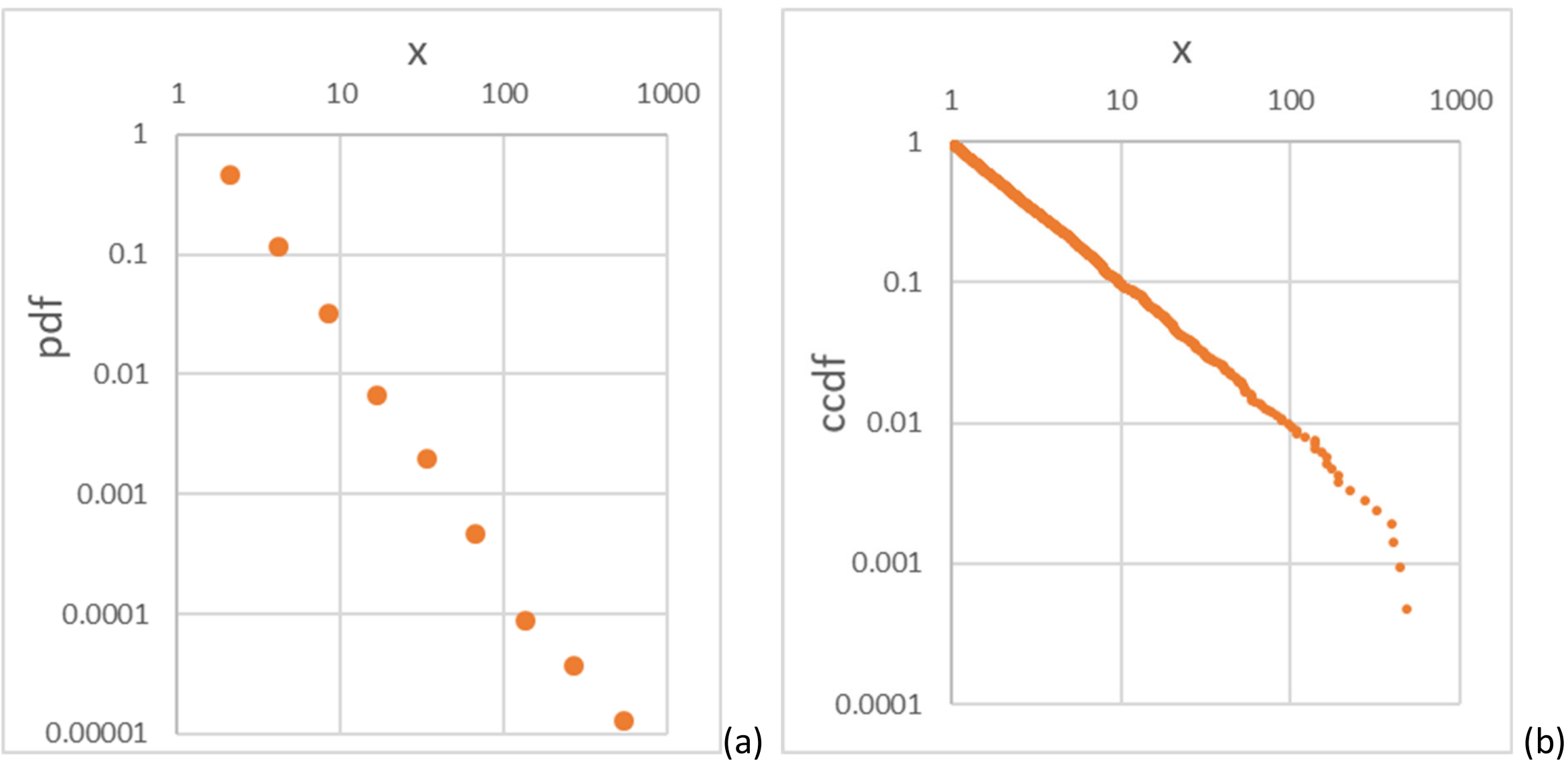

*Figure A3. PDF (a) and CCDF (b) of two thousand data points following a power law distribution $p(x) = x^{-2}$ for x > 1.*

Both look more or less like straight lines on the logarithmic scale just as they should. Now let us truncate the distribution. We removed from the sample all the data points which are less than 2 (there were 990 of those) or more than 50 (there were 40 of those). Figure A4 shows the binned distribution of the remaining 970 data points (a) and the CCDF (b).

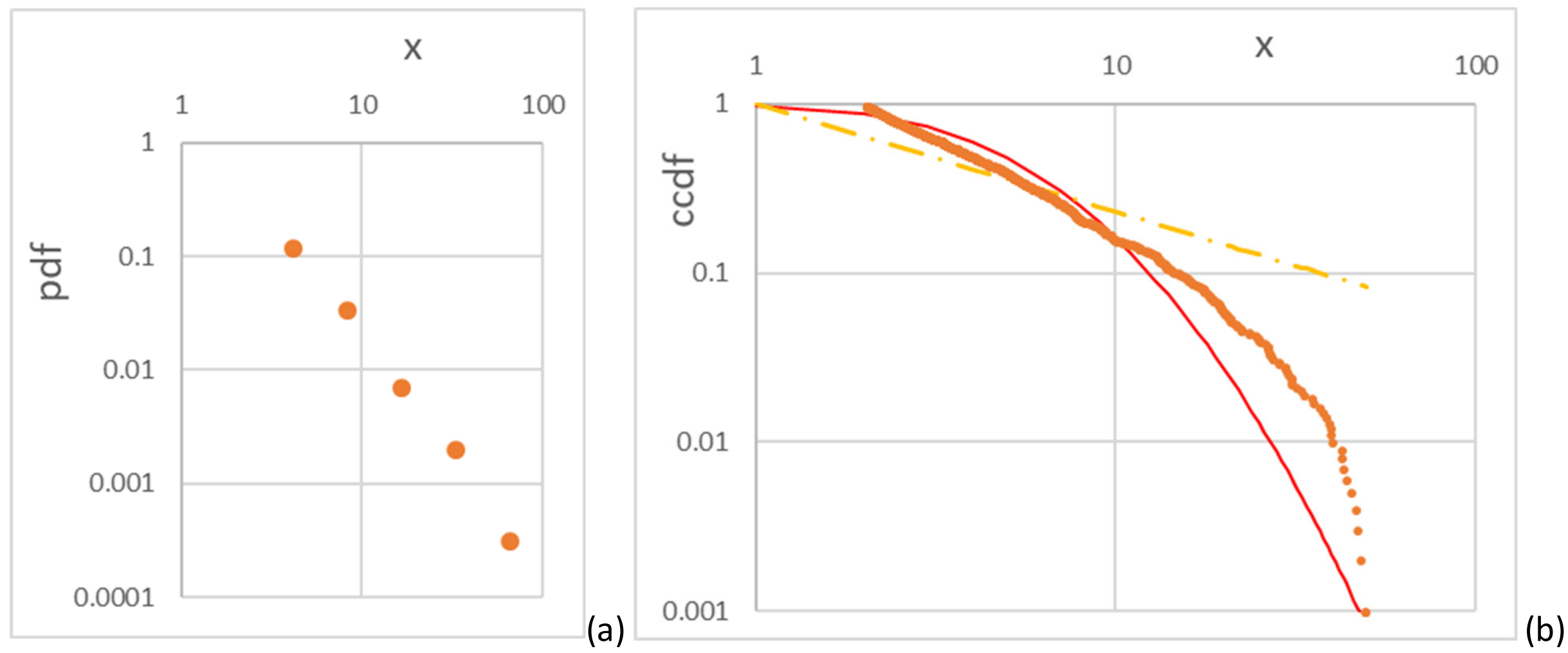

*Figure A4. Same as Figure A3 but for the truncated sample.*

The binned plot shows same good distribution, only fewer data points. The CCDF is clearly not a power law. In fact, a lognormal distribution (solid line) fits better than a power law (dot-dashed line) as one can see in Figure A4 (b). But there was nothing lognormal there in principle, for the distribution is a truncated power law. To understand what happened let us recall the definition of the CCDF. If we have $N$ data points $x_1 < x_2 < x_3 < \ldots < x_N$ then

$$CDF(x_i) = i / N \quad ; \quad CCDF(x_i) = 1 - CDF(x_i) = (N - i) / N$$

We have $CCDF(x_N) = (N-N)/N = 0$ so the last data point does not show up on the logarithmic plot. If we remove $K$ largest data points and compute the new CCDF for the remaining set $x_1 < x_2 < x_3 < \ldots < x_{N-K}$ we get

$$CCDF_1(x_i) = 1 - CDF_1(x_i) = (N - K - i) / (N - K)$$

Now let us count the data points from the right: $i = N - K - m$. This way $m = 1$ corresponds to the rightmost data point showing up on the plot of CCDF for the shrunk data set, $m = 2$ corresponds to the second from the right data point and so on. We get:

$$CCDF_1(x_i) = m / (N - K) ; \quad CCDF(x_i) = (K + m) / N$$

And for the ratio:

$$CCDF_1(x_i) / CCDF(x_i) = m / (K + m) * (N / (N - K))$$

The factor $N / (N - K)$ does not depend on $m$ and merely rescales the CCDF. The factor m / (K + m) varies a lot. In the Figure A4 we have $K=40$ so for $m=1$ the factor is 1/41. So, the point $m = 1$ drops forty times below the line. This is exactly what we see in the figure. When $m >> K$ the factor becomes 1. It reaches 0.9 (10% below the limit value) when $m = 9K$. This means that removing 40 data points distorts 360 of the remaining data points by more than 10%. This is why we get the huge distortion of the power law.

It is easy to see how to fix this problem for the shrunk data set. One has to compute $CCDF_1$ using $N$ instead of $N - K$. The result is in Fig. A5.

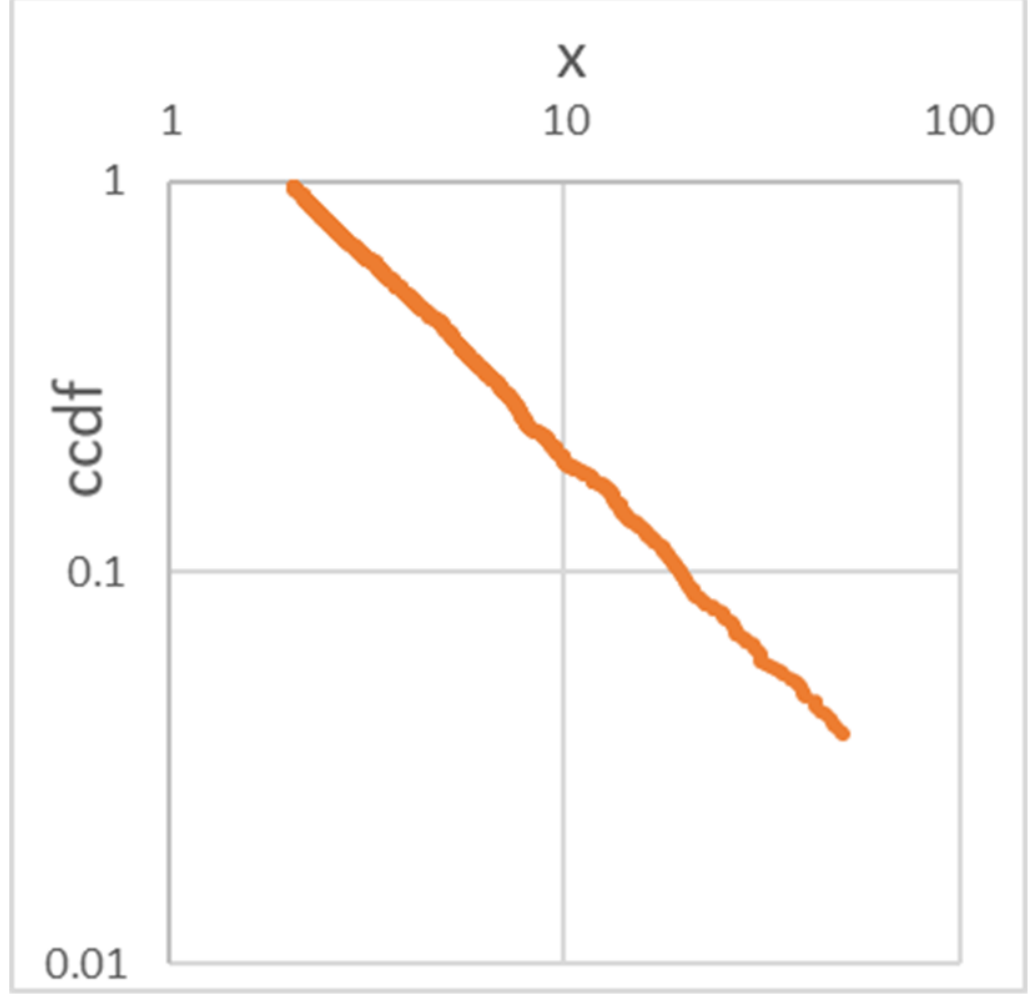

*Figure A5. Fixed CCDF for the truncated distribution.*

This suggests a way to fix the CCDF for the inter-murder intervals. This distribution is also truncated, not by hand like in the toy model we just described, but in a natural way by the limit of human active life span. So, we need to plot it as if in addition to 607 actual intervals there were $Z$ imaginary intervals. To find $Z$ we can use the power law fit in Fig. A2. In the region between 64 days and 8192 days the least square fit gives $y = 4.6 * x^{-1.6}$. The bins above 8192 are empty. However, we imagine that this power law continues to infinity and estimate: $Z = 607 * 4.6 * \int_{8192}^{\infty} x^{-1.6} dx \approx 21$. Now we plot the CCDF as if above the largest inter-murder interval there were 21 more intervals (Fig. A6). The least square fit for the intervals of 33 or more days (should do like that because the 64 days bin in PDF includes all intervals between 33 and 64) produces a power law with the exponent 0.55. This is in good agreement with the 1.6 exponent obtained for the binned PDF by the least-square fit. A maximum likelihood fit for the intervals of 33 or more days gives a very close exponent of 1.63.

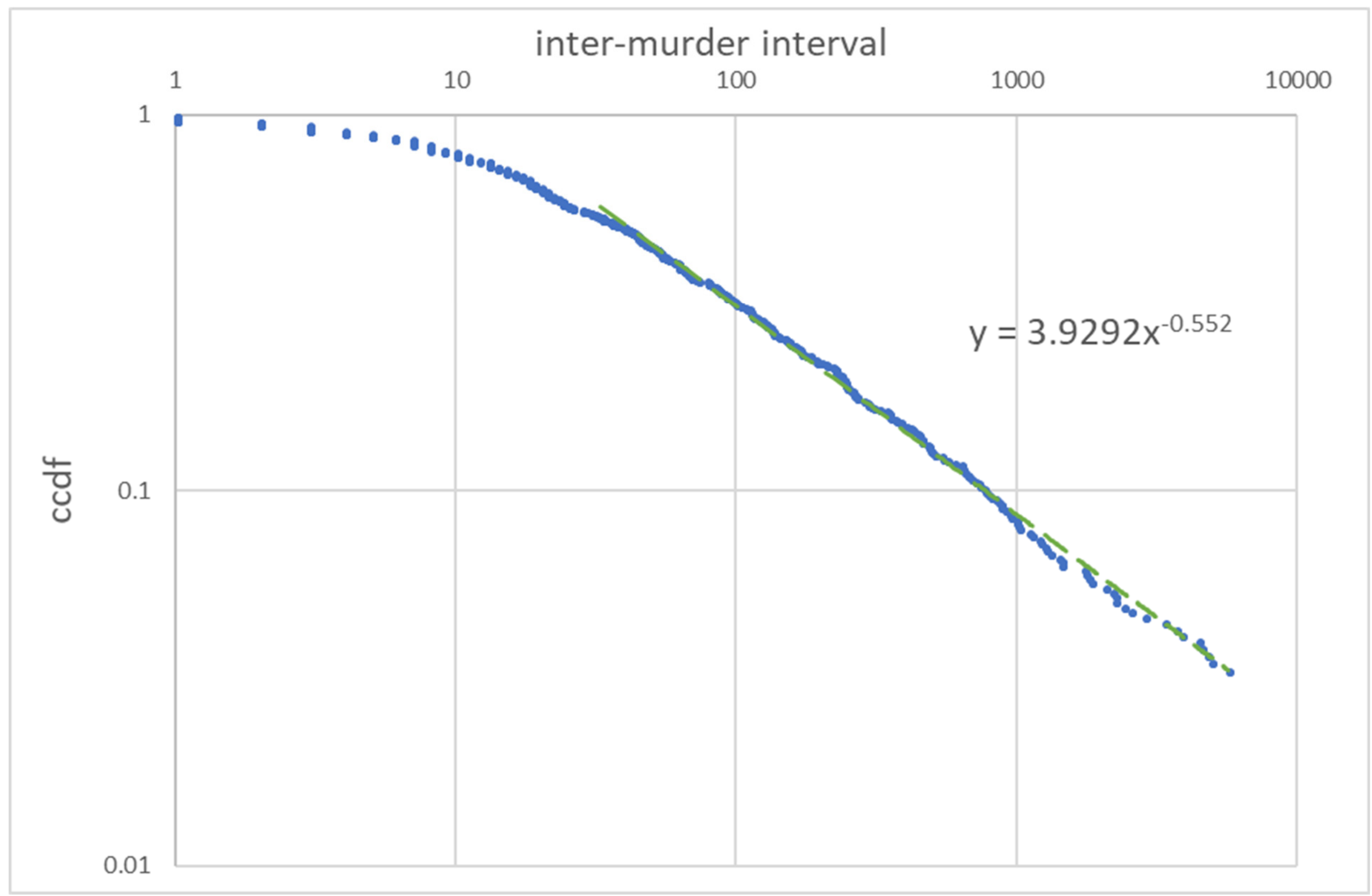

*Figure A6. Fixed CCDF for inter-murder intervals.*